\def \defvmargin{2.5cm}
\def \defhmargin{1.5cm}

\documentclass[12pt]{article}
\input{my-main-template}

\newcommand{\ie}{i.e.\xspace}
\newcommand{\eg}{e.g.\xspace}
\newcommand{\eee}{e}

\allowdisplaybreaks

\begin{document}

\title{Optimality of Correlated Sampling Strategies}

\author{\setlength\tabcolsep{.7em}\def\arraystretch{1.5}
\begin{tabular}{ccc}
Mohammad Bavarian\thanks{Department of Mathematics and CSAIL, MIT. Supported in part by NSF Award CCF-1420692. {\tt bavarian@mit.edu}.} &
Badih Ghazi\thanks{CSAIL, MIT. Supported in part by NSF CCF-1420956, NSF CCF-1420692 and CCF-1217423. {\tt badih@mit.edu}.} &
Elad Haramaty\thanks{Harvard John A. Paulson School of Engineering and Applied Sciences. Part of this work supported by NSF Award CCF-1565641.
{\tt seladh@gmail.com}.}\\
Pritish Kamath\thanks{CSAIL, MIT. Supported in part by NSF CCF-1420956 and NSF CCF-1420692.  {\tt pritish@mit.edu}.} &
Ronald L. Rivest\thanks{Institute Professor, MIT. This work supported by the Center for Science of Information (CSoI), an NSF Science and Technology Center, under grant agreement CCF-0939370. {\tt rivest@mit.edu}} &
Madhu Sudan\thanks{Harvard John A. Paulson School of Engineering and Applied Sciences. Part of this work supported by NSF Award CCF-1565641 and a Simons Investigator Award. {\tt madhu@cs.harvard.edu}.}
\end{tabular}
}

\maketitle

\begin{abstract}
In the \emph{correlated sampling} problem, two players
are given probability distributions $P$ and $Q$, respectively,
over the same finite set,
with access to shared randomness.
Without any communication, the two players are each required to output an
element sampled according to their respective distributions, while trying to
minimize the probability that their outputs disagree.
A well known strategy due to Kleinberg--Tardos and Holenstein, with a close variant (for a similar problem) due to Broder, solves this task with disagreement probability at most $2 \delta/(1+\delta)$, where $\delta$ is the total variation distance between $P$ and $Q$. This strategy has been used in several different contexts, including sketching algorithms, approximation algorithms based on rounding linear programming relaxations, the study of parallel repetition and cryptography.

In this paper, we give a surprisingly simple proof that this strategy is
essentially optimal. Specifically, for every $\delta \in (0,1)$,
we show that any correlated sampling strategy incurs a
disagreement probability of essentially $2\delta/(1+\delta)$ on some inputs $P$ and $Q$ with total variation distance at most $\delta$. This partially answers a recent question of Rivest.

Our proof is based on studying a new problem that we call \emph{constrained agreement}. Here, the two players are given subsets $A \subseteq [n]$ and $B \subseteq [n]$, respectively, 
and their goal is to output an element $i \in A$ and $j \in B$, respectively, 
while minimizing the probability that $i \neq j$. We prove tight bounds for this question, which in turn imply tight bounds for correlated sampling. Though we settle basic questions about the two problems, our formulation leads to more
fine-grained questions that remain open.

\end{abstract}

\newpage

\section{Introduction}
In this paper,  
we study \emph{correlated sampling}, a
basic task, variants of which have been considered in the context of sketching algorithms~\cite{broder1997resemblance}, approximation algorithms based on rounding linear programming relaxations \cite{kleinberg2002approximation, charikar2002similarity}, the study of parallel repetition \cite{holenstein2007parallel, rao2011parallel, barak2008rounding} and cryptography~\cite{Rivest}. 

The \emph{correlated sampling problem} is defined as 
follows. Alice and Bob are given probability distributions $P$ and $Q$,
respectively, over a finite 
set $\Omega$.   
Without any communication, using only shared randomness as the means to coordinate, Alice is required to output an element $a$ distributed according to $P$ and Bob is required to output an element $b$ distributed according to $Q$. Their goal is to minimize the disagreement probability $\Pr[a \neq b]$, which we will compare
with  
the total variation distance between $P$ and $Q$, defined as
\begin{equation}
	\dTV(P,Q) ~:=~ \sup_{A \subseteq \Omega} P(A) - Q(A) ~=~ \frac{1}{2} \sum_{\omega \in \Omega} |P(\omega) - Q(\omega)|.
\end{equation}
A \emph{correlated sampling strategy} is formally defined below, where $\Delta_{\Omega}$ denotes the set of all probability distributions over $\Omega$ and $(\calR, \calF, \mu)$ denotes the probability space corresponding to the randomness shared by Alice and Bob. $\calR$ is the sample space, $\calF$ is a $\sigma$-algebra over $\calR$ and $\mu$ is a probability measure over $(\calR, \calF)$. Even though $\Omega$ is finite, we allow $\calR$ to be infinite. For simplicitly, we abuse notation and use $\calR$ to denote both the sample space and the probability space.

\begin{definition}\label{def:corr_samp}
	A \emph{correlated sampling strategy} for a finite
	set $\Omega$   
	with error $\eps : [0,1] \to [0,1]$ is specified by a probability space
	$\calR$ and a pair of 
	functions\footnote{We require both functions to be measurable in their second argument.} 
	$f, g : \Delta_\Omega \times \calR \to \Omega$,
	such that for all $P, Q \in \Delta_\Omega$ with $\dTV(P,Q) \le \delta$,
	the following hold.  
	\begin{center} {\renewcommand{\arraystretch}{1.3}
			\begin{tabular}{ll}
				{\bf [Correctness]} & $\set{f(P,r)}_{r \sim \calR} = P$ and $\set{g(Q,r)}_{r \sim \calR} = Q$,\\
				{\bf [Error Guarantee]} & $\Pr_{r \sim \calR} \insquare{f(P,r) \ne g(Q,r)} ~\le~ \eps(\delta)$.
		\end{tabular}}
	\end{center}
\end{definition}
\noindent In the above, $\set{f(P,r)}_{r \sim \calR}$ denotes the push-forward measure, that is, the distribution of the random variable $f(P,r)$ over $\Omega$, where $r \sim \calR$ is the source of shared randomness. 
For simplicity, we will often not mention $\calR$ explicitly when talking about correlated sampling strategies.  While we defined correlated sampling strategies for finite 
sets 
only, it is possible to define it for infinite
sets $\Omega$;   
see \Cref{sec:discussions} for a discussion.
In this paper we consider finite sets $\Omega$ only, except where 
otherwise stated.

A correlated sampling strategy is notably different from the fundamental notion of a \emph{coupling} (see, \eg,~\cite{thorisson2000coupling} for an introduction), where we require a \emph{single} coupling function $h : \Delta_{\Omega} \times \Delta_{\Omega} \to \Delta_{\Omega \times \Omega}$ such that for any distributions $P$ and $Q$ it holds that the marginals of $h(P,Q)$ are $P$ and $Q$ respectively. In other words, a coupling function has the knowledge of both $P$ and $Q$, whereas a correlated sampling strategy operates locally 
on the knowledge of $P$ and on the knowledge of $Q$. 
It is well known that for any coupling function $h$, it holds that $\Pr_{(a,b)\sim h(P,Q)}[a \ne b] \ge \dTV(P,Q)$ and that this bound is achievable. Observe that a correlated sampling strategy induces a coupling given as $\set{(f(P,r),g(Q,r))}_{r \sim \calR}$. Thus, it follows that $\eps(\delta) \ge \delta$. And yet a priori, it is unclear whether any non-trivial correlated sampling strategy can even exist, since the error $\eps$ is not allowed to depend on the size of $\Omega$.

Somewhat surprisingly, there exists a simple strategy whose error can be bounded by roughly twice the total variation distance (and in particular does not degrade with the size of $\Omega$). Variants of this strategy have been rediscovered multiple times in the literature, 
yielding the following theorem.

\begin{theorem}[\cite{broder1997resemblance, kleinberg2002approximation, holenstein2007parallel}]\label{thm:corr-samp}
	For all $n \in \bbZ_{\ge 0}$, there exists a correlated sampling strategy for sets of size $n$, with error $\eps : [0,1] \to [0,1]$, such that for all $\delta \in [0,1]$, it holds that
	\begin{equation}\label{eq:Holest_bd}
		\eps(\delta) ~\le~ \frac{2 \cdot \delta}{1+\delta}.
	\end{equation}
\end{theorem}

Strictly speaking,
Broder's paper~\cite{broder1997resemblance} did
not consider the general correlated sampling problem. Rather it introduced the 
\emph{MinHash strategy}, which can be
interpreted    
as a correlated sampling strategy for the special case where $P$ and $Q$ are \emph{flat} distributions, \ie, they are uniform over some subsets of $\Omega$. In particular, if $P = \calU(A)$ and $Q = \calU(B)$ are distributions that are uniform over sets $A,\, B \subseteq \Omega$, respectively, 
then the MinHash strategy gives an error probability of
$1 - |A \cap B|/|A \cup B|$,
also known as the \emph{Jaccard distance} between $A$ and $B$.
In the special case when $|A| = |B|$, this is equivalent to the bound above.

The technique can be generalized to other (non-flat) distributions to get the bound in \autoref{thm:corr-samp}, thereby yielding a strategy due to Kleinberg--Tardos and
Holenstein.\footnote{Strictly speaking,   
	if $P$ and $Q$ are flat over
	subsets of different sizes,  
	the above bound is weaker than that obtained from a direct application of the MinHash strategy. See \Cref{sec:discussions} for a discussion.}
Several variants of this (sometimes referred to as ``consistent sampling'' protocols) have been used in
applied work, including
\cite{manber1994finding,gollapudi2006dictionary, manasse2010consistent, haeupler2014consistent}.

Given \autoref{thm:corr-samp}, a natural and basic question is whether the bound on the error can be improved; the only lower bound we are aware of is the one that arises from coupling, namely $\eps(\delta) \ge \delta$. This question was 
recently raised by Rivest~\cite{Rivest} in the context of a new encryption scheme and was one of the motivations for this work. We give a surprisingly simple proof that the bound in \autoref{thm:corr-samp} is essentially tight.


\begin{theorem}[Main Result]\label{thm:main}
	For all $\delta, \gamma \in (0,1)$, for all sufficiently large $n$, any correlated sampling strategy for a set of size $n$ with error $\eps: [0,1] \to [0,1]$ satisfies
	\begin{equation}\label{eq:main_bd}
		\eps(\delta) ~\ge~ \frac{2 \cdot \delta}{1+\delta} - \gamma.
	\end{equation}
\end{theorem}

\noindent {\bf Organization of the paper.} In \Cref{sec:pf_main}, we prove \autoref{thm:main}. In \Cref{sec:special_cases}, we consider the setting where $\Omega$ is of a fixed finite size, which was the question originally posed by Rivest~\cite{Rivest}. In this regime, there turns out to be a surprising strategy that gets better error than \autoref{thm:corr-samp} in a very special case.
However, it was conjectured in~\cite{Rivest} that in fact a statement like \autoref{thm:main} holds in every other case and we make progress on this conjecture by proving it in one such case.  For completeness, 
in \Cref{sec:protocols} we describe  
the correlated sampling strategies of Broder, Kleinberg--Tardos,
and Holenstein, underlying \autoref{thm:corr-samp}.
We conclude with some more observations and open questions
in \Cref{sec:discussions}.  


\paragraph{Acknowledgements}
We thank anonymous ToC reviewers for their feedback that has significantly helped improve the presentation of this paper. We are also grateful to Laci Babai for detailed comments that helped reorganize the paper and brought in more clarity to the discussion about correlated sampling for infinite spaces.

\section{Lower bound on correlated sampling}\label{sec:pf_main}

\noindent In order to prove \autoref{thm:main}, we first introduce the
\emph{constrained agreement} problem, a  
relaxation of the correlated sampling problem.  In this problem,  
Alice and Bob are given sets $A \subseteq \Omega$ and $B \subseteq \Omega$,
respectively,
where the pair $(A,B)$ is sampled from some (known) distribution $\mathcal{D}$.
Alice and Bob are required to output elements $a \in A$ and $b \in B$,
respectively, such that the disagreement probability $\Pr_{(A,B) \sim D}[a \neq b]$ is minimized.

This can be viewed as a relaxation of the correlated sampling problem by first considering the case of \emph{flat} distributions in \autoref{def:corr_samp} and relaxing the restrictions of $\set{f(P,r)}_{r \sim \calR} = P$ and $\set{g(Q,r)}_{r \sim \calR} = Q$ to only requiring that $f(P,r) \in \supp(P)$ and $g(Q,r) \in \supp(Q)$ for all $r \in \calR$. This makes it a constraint satisfaction problem and we consider a distributional version of the same.

In the following definition, we use $2^{\Omega}$ to denote the powerset of $\Omega$.

\begin{definition}\label{def:const_ag}
	A \emph{constrained agreement strategy} for a finite
	set 
	$\Omega$ and a probability distribution $\calD$ over $2^\Omega \times 2^\Omega$ is specified by a pair of functions $f, g : 2^\Omega \to \Omega$ and has error $\mathrm{err}_{\calD}(f,g)$ if the following hold.  
	\begin{center} {\renewcommand{\arraystretch}{1.3}
			\begin{tabular}{ll}
				{\bf [Correctness]} & $\forall A, B \subseteq \Omega\; : \  f(A) \in A$ and $g(B) \in B$,\\
				{\bf [Error guarantee]} & $\Pr_{(A,B) \sim \calD} \insquare{f(A) \ne g(B)} ~=:~ \mathrm{err}_{\calD}(f,g)$.
		\end{tabular}}
	\end{center}
\end{definition}

Note that since the constrained agreement problem is defined with respect to a (known) probability distribution $\mathcal{D}$ on pairs of sets, we can require,
without loss of generality, that the strategies $(f,g)$ be deterministic (since any randomized strategy can be derandomized with no degradation in the error).

In order to prove \autoref{thm:main}, we characterize the optimal constrained agreement strategy in the special case when $\calD = \calD_p$ where every element $\omega \in \Omega$ is independently included in each of $A$ and $B$ with probability $p$.

\begin{lemma}\label{le:prod_opt}
	For all $p \in [0,1]$, any constrained agreement strategy $(f,g)$ for a finite
	set 
	$\Omega$ and distribution $\calD = \calD_p$ over $2^{\Omega} \times 2^{\Omega}$, has error
	\[
	\mathrm{err}_{\calD_p}(f,g) \ge \frac{2(1-p)}{2-p}.
	\]
\end{lemma}
\begin{proof}[Proof of \autoref{le:prod_opt}]
	For ease of notation, let $\Omega = [n]$.
	Let $(f,g)$ be a constrained agreement strategy. We will construct functions $f^*$ and $g^*$ such that
	\[
	\mathrm{err}_{\calD_p}(f,g) ~\ge~ \mathrm{err}_{\calD_p}(f^*,g^*) ~\ge~ \frac{2(1-p)}{2-p}.
	\]
	
	For every $i \in [n]$, let $\beta_i := \Pr_{B}[g(B) = i]$. Without loss of generality (by suitably permuting $[n]$), we can assume that $\beta_{1} \geq \beta_{2} \geq \dots \geq \beta_{n}$. Since $A$ and $B$ are independently sampled in $\mathcal{D}_p$, it follows that when Bob's strategy is fixed to $g$, the strategy of Alice that results in the largest agreement probability is simply $f^*(A) := \argmax_{i \in A} \beta_i = \min\set{i : i \in A}$ for all $A \subseteq [n]$.
	
	So far we have $\mathrm{err}_{\calD_p}(f,g) \ge \mathrm{err}_{\calD_p}(f^*,g)$.
	We can repeat the same process again. For every $i \in [n]$, define $\alpha_i := \Pr_{A}[f^*(A) = i]$. Due to the specific choice of $f^*$, it holds that $\alpha_i = (1-p)^{i-1} p$ and hence $\alpha_1 \ge \alpha_2 \ge \cdots \ge \alpha_n$. Thus, when Alice's strategy is fixed to $f^*$, the strategy of Bob that results in the largest agreement probability is given by $g^*(B) = \argmax_{i \in B} \alpha_i = \min\set{i : i \in B}$ for all $B \subseteq [n]$.
	Hence, 
	we get $\mathrm{err}_{\calD_p}(f,g) \ge \mathrm{err}_{\calD_p}(f^*,g) \ge \mathrm{err}_{\calD_p}(f^*,g^*)$ where
	\begin{align*}
		\mathrm{err}_{\calD_p}(f^*,g^*) &~:=~ 1 - \Pr_{(A,B) \sim \mathcal{D}_p}[f^*(A) = g^*(B)]\\ 
		&~=~ 1 - \displaystyle\sum\limits_{i=1}^n \Pr_{A}[f^*(A) = i] \cdot \Pr_{B}[g^*(B) = i]\\ 
		&~=~ 1- \displaystyle\sum\limits_{i=1}^n (1-p)^{2 \cdot (i-1)} \cdot p^2\\ 
		&~\ge~ 1 - \frac{p}{2-p} ~=~ \frac{2(1-p)}{2-p}.
	\end{align*}
	Thus, we conclude that
	\[
	\mathrm{err}_{D_p}(f,g) ~\ge~\frac{2(1-p)}{2-p}.\qedhere
	\]
\end{proof}

\noindent Before turning to the proof of \autoref{thm:main}, we note a couple of basic facts.
\begin{fact}\label{fact:tv-flat}
	For \emph{flat} distributions $P = \calU(A)$ and $Q = \calU(B)$ with $A, B \subseteq \Omega$, it holds, that
	$$\dTV(P,Q) ~=~ 1 - \frac{|A \cap B|}{\max\set{|A|,|B|}}.$$
\end{fact}

\noindent The following concentration bound, due to Sergei Bernstein from the 1920s,
is often referred to as Chernoff's or Hoeffding's bound. The Bernoulli
random variable    
$X \sim \mathrm{Ber}(p)$ is a $0$-$1$ random variable with $Pr[X=1] = p$. 
\begin{fact}[See, \eg,   
	Cor~4.6 in~\cite{mitzenmacher2005probability})]\label{fact:hoeffding}
	For $X_1, \ldots, X_n$ drawn i.\,i.\,d.\ from $\mathrm{Ber}(p)$, it holds for all $\tau > 0$, that
	$$ \Pr\insquare{\inabs{\sum_i X_i - pn} \ge \tau \cdot pn} \le 2 \cdot \eee^{-pn\tau^2/3}.$$
\end{fact}

\begin{proof}[Proof of \autoref{thm:main}]
	Fix $\delta, \gamma \in (0,1)$. Assume, for the sake of contradiction, that for infinitely many values of $n$, there is a correlated sampling strategy $(f^*,g^*)$ for a set of size $n$ with error
	\[
	\eps(\delta) < \frac{2 \cdot \delta}{1+\delta} - \gamma.
	\]
	Let $\delta' \in (0,\delta)$ be such that
	\begin{equation}\label{eq:delta_prime}
		\frac{2 \cdot \delta}{1+\delta} - \gamma < \frac{2 \cdot \delta'}{1+\delta'} < \frac{2 \cdot \delta}{1+\delta}.
	\end{equation}
	Consider the distribution $\mathcal{D}_p$ over pairs $(A,B)$ of subsets $A,B \subseteq [n]$ where each $i \in [n]$ is independently included in each of $A$ and $B$ with probability $p := 1-\delta'$. Thus, we have $\Ex[|A|] = \Ex[|B|] = p \cdot n$, and $\Ex[|A \cap B|] = p^2 \cdot n$. Moreover, by \autoref{fact:hoeffding} with $\tau=n^{-0.2}$,  
	we have that
	\begin{align*}
		\Pr_{A}[||A| - pn| > p \cdot n^{0.8}] &~\le~ 2 \cdot \eee^{-p \cdot n^{0.6} /3},\\
		\Pr_{B}[||B| - pn| > p \cdot n^{0.8}] &~\le~ 2 \cdot \eee^{-p \cdot n^{0.6} /3},\\
		\Pr_{A,B}[||A \cap B| - p^2 n| > p^2 \cdot n^{0.8}] &~\le~ 2 \cdot \eee^{-p^2 \cdot n^{0.6} /3}.
	\end{align*}
	Hence, by the union bound and using $p^2 \le p$, we get that with probability at least $1 - 6\cdot \eee^{-p^2 \cdot n^{0.6} /3}$, we have that $||A| - p\cdot n| \le p n^{0.8}$, $||B| - p\cdot n| \le p n^{0.8}$ and $||A \cap B| - p^2\cdot n| \le p^2 n^{0.8}$. Thus, for the distributions $P = \mathcal{U}(A)$ and $Q = \mathcal{U}(B)$, it holds with probability at least $1 - 6\cdot \eee^{-p^2 \cdot n^{0.6} /3}$ that
	\begin{eqnarray*}
		\dTV(P,Q) &=& 1 - \frac{|A \cap B|}{\max\{|A|,|B|\}}\\ 
		&\le& 1-p + o_n(1)\\ 
		&=& \delta' + o_n(1)\\
		&<& \delta \quad \text{for sufficiently large $n$.}
	\end{eqnarray*}
	The assumed strategy $(f^*, g^*)$ is such that
	\[
	\Pr_{r \sim \calR} [f(P,r) \ne g(Q,r)] \le \frac{2\delta}{(1+\delta)} - \gamma
	\]
	when $\dTV(P,Q) \le \delta$ and at most $1$ otherwise. In our random
	choice of the pair of distributions $(P,Q)$, the probability of
	$\dTV(P,Q) > \delta$ is at most $o_n(1)$. Thus,
	\[
	\Pr_{(P,Q),\, r \sim \calR} [f(P,r) \ne g(Q,r)] \le \frac{2 \cdot
		\delta}{1+\delta} - \gamma + o_n(1)
	\]
	when applied on the randomly
	sampled $(P,Q)$. In particular, by averaging, there exists a
	deterministic constrained agreement strategy with no worse
	disagreement probability. That is,
	\begin{equation}\label{eq:min_ub}
		\exists \ (f, g), \quad \mathrm{err}_{\calD_p}(f,g) ~\le~ \frac{2 \cdot \delta}{1+\delta} - \gamma + o_n(1).
	\end{equation}
	But from \autoref{le:prod_opt} we have that,
	\begin{equation}\label{eq:min_lb}
		\forall \ (f, g), \quad \mathrm{err}_{\calD_p}(f,g) ~\ge~ \frac{2(1-p)}{2-p} = \frac{2 \cdot \delta'}{1+\delta'}.
	\end{equation}
	Putting \eqref{eq:min_ub} and \eqref{eq:min_lb} together contradicts \eqref{eq:delta_prime} for sufficiently large $n$.
\end{proof}

\section{Correlated sampling over a fixed set of finite size}\label{sec:special_cases}
\autoref{thm:main} establishes that the correlated sampling strategy underlying \autoref{thm:corr-samp} is \emph{nearly} optimal for $\Omega$ that is sufficiently large in size. However, it does not say that the strategy underlying \autoref{thm:corr-samp} is exactly optimal for a fixed set of finite size. The quest for understanding optimality in this setting was motivated by a new encryption scheme proposed by Rivest~\cite{Rivest}. But as we will see shortly, this quest is not entirely straightforward!

In order to elaborate on this, it will be useful to formally define restricted versions of the correlated sampling strategy which are required to work only when the input pair $(P,Q)$ is promised to lie in a given relation $\calG \subseteq \Delta_{\Omega} \times \Delta_{\Omega}$.

\begin{definition}
	For a finite
	set 
	$\Omega$ and a relation $\calG \subseteq \Delta_{\Omega} \times \Delta_{\Omega}$, a \emph{$\calG$-restricted correlated sampling strategy} with error $\eps$ is specified by a probability space $\calR$, a pair of functions $f, g : \Delta_\Omega \times \calR \to \Omega$ if the following hold for all pairs of distributions $(P, Q) \in \calG$,
	\begin{center} {\renewcommand{\arraystretch}{1.3}
			\begin{tabular}{ll}
				{\bf [Correctness]} & $\set{f(P,r)}_{r \sim \calR} = P$ and $\set{g(Q,r)}_{r \sim \calR} = Q$,\\
				{\bf [Error Guarantee]} & $\Pr_{r \sim \calR} \insquare{f(P,r) \ne g(Q,r)} ~\le~ \eps$.
		\end{tabular}}
	\end{center}
\end{definition}

\noindent For example, letting $\calG$ to be set of all pairs $(P,Q)$ with $\dTV(P,Q) \le \delta$ essentially recovers the original setting of correlated sampling, for a fixed total variation distance bound between the input distributions. For the rest of this section, we will consider a special kind of $\calG$-restriction corresponding to Alice and Bob having
\emph{flat distributions} over $\Omega = [n]$.   

\begin{definition}
	For all $n$, the relation $\calG^n_{a,b,\ell} \subseteq \Delta_{[n]} \times \Delta_{[n]}$ is defined to consist of
	pairs $(P,Q)$ of flat distributions   
	corresponding to sets $A, B \subseteq [n]$ such that $P = \calU(A)$, $Q = \calU(B)$ and $|A| = a$, $|B| = b$, $|A \cap B| = \ell$.
	(For the relation to be non-empty, it is required that $\ell \le \min\set{a, b}$ and $a + b - \ell \le n$.)
\end{definition}
\noindent Recall from \autoref{fact:tv-flat}, that for all $(P,Q) \in \calG^n_{a,b,\ell}$ with $P = \calU(A)$ and $Q = \calU(B)$, is given by
\[ \dTV(P,Q) ~=~ 1 - \frac{|A \cap B|}{\max \set{|A|, |B|}} ~=~ 1 - \frac{\ell}{\max \set{a, b}}. \]
Moreover, the MinHash strategy applied on input pairs $(P, Q) \in \calG^n_{a,b,\ell}$ has a disgreement probability
\[
1 - \frac{|A \cap B|}{|A \cup B|} ~=~ 1 - \frac{\ell}{a + b - \ell}.
\]
One might suspect that this is optimal for all values of $n$, $a$, $b$ and $\ell$. But rather surprisingly, in the very special case where $|A \cap B| = 1$ and $|A \cup B| = n$, Rivest~\cite{Rivest} gave a strategy with smaller error probability than the above! While we don't know of any applications for this strategy itself, its purpose here is to illustrate that there can be strategies which do better than the MinHash strategy in some special cases.

\begin{theorem}[\cite{Rivest}]\label{thm:rivest}
	For all $a, b \in \bbZ_{\ge 1}$ there exists a $\calG^{a+b-1}_{a,b,1}$-restricted correlated sampling strategy with error at most $1 - 1/\max\set{a,b}$.
\end{theorem}
\noindent For completeness, we describe this strategy in \Cref{sec:special_protocols}. Note that for $(P, Q) \in \calG^{a+b-1}_{a,b,1}$,
\begin{equation*}
	\dTV(P,Q) ~=~ 1 - \frac{1}{\max\set{a,b}} ~<~ 1 - \frac{1}{a+b-1}.
\end{equation*}
This naturally leads to the question: \emph{Is there a better correlated sampling strategy for larger intersection sizes?} In fact, the MinHash strategy was conjectured to be optimal 
in every other case (in particular, for all $\ell > 1$) 
by Rivest~\cite{Rivest} as this is sufficient for proving 
the security of his proposed encryption scheme.

\begin{conjecture}[Rivest] 
	\label{conj:rivest}
	For every collection of positive integers $n \ge a,b \ge \ell \ge 2$ such that $n \geq a+b - \ell$, any $\calG^{n}_{a,b,\ell}$-restricted correlated sampling strategy makes error at least $1 - \ell/(a+b-\ell)$.
\end{conjecture}

\noindent As partial progress towards this conjecture, we prove that in the other extreme (as compared to \autoref{thm:rivest}), the above conjecture does hold. In particular, we show the following theorem in \Cref{sec:large_int_pf}.

\begin{theorem}\label{thm:opt_large_inter}
	For all $a=b \ge 1$, $\ell = a-1$ and $n \ge a + 1$, any $\calG^{n}_{a,b,\ell}$-restricted correlated sampling strategy makes error at least $1 - \ell/(a+b-\ell)$.
\end{theorem}


\subsection{Correlated sampling strategy of Rivest~\cite{Rivest}}\label{sec:special_protocols}

In order to prove \autoref{thm:rivest}, we recall
Philip Hall's ``Marriage Theorem.''  

\begin{lemma}[P. Hall; see, \eg,~\cite{van2001course}]  
	\label{th:hall} 
	Fix a
	bipartite graph $G$ on vertex sets $L$ and $R$ (with $|L| \le |R|$). There exists a matching that entirely covers $L$ if and only if for every subset $S \subseteq L$, we have that $|S| \le |N_G(S)|$, where $N_G(S)$ denotes the set of neighbors in $G$ of vertices in $S$.
\end{lemma}

\begin{proof}[Proof of \autoref{thm:rivest}]
	First, let us consider the case where $a = b$. Let $\binom{[n]}{a}$ denote the set of all subsets $A \subseteq [n]$ with $|A| = a$. Consider the bipartite graph $G$ on vertices $\binom{[n]}{a} \times \binom{[n]}{a}$, with an edge between vertices $A$ and $B$ if $|A \cap B| = 1$. It is easy to see that $G$ is $a$-regular (since $n = 2a-1$). Iteratively using \autoref{th:hall}, we get that the edges of $G$ can be written as a disjoint union of $a$ matchings. Let us denote these as $M_1, M_2, \ldots, M_a$.
	
	The $\calG^{2a-1}_{a,a,1}$-restricted correlated sampling strategy of Alice and Bob is as follows: Use the shared randomness to sample a random index $r \in [a]$ and consider the matching $M_r$. If $(A,B')$ is the edge present in $M_r$, then Alice outputs the unique element in $A \cap B'$. Similarly, if $(A',B)$ is the edge present in $M_r$, then Bob outputs the unique element in $A' \cap B$. This strategy is summarized in \Cref{prot:rivest}.
	
	\begin{figure}[ht]
		\centering
		\begin{minipage}{0.95\linewidth}
			\begin{algorithm}[H]
				{\bf Alice's input:} $A \subseteq [n]$\\[1mm]
				{\bf Bob's input:} $B \subseteq [n]$\\[1mm]
				{\bf $\calG$-restriction:} $|A| = |B| = a$, $|A \cap B| = 1$ and $A \cup B = [n]$, \ie, $n = a+b-1$\\[1mm]  
				{\bf Preprocessing:}  
				Let $G$ be the bipartite graph on vertices $\binom{[n]}{a} \times \binom{[n]}{a}$, with an edge between vertices $A$ and $B$ if $|A \cap B| = 1$. Decompose the edges of $G$ into $a$ disjoint matchings $M_1, \ldots, M_a$.\\[1mm]
				{\bf Shared randomness:} Index $r \sim \calU([a])$\\[2mm]
				{\bf Strategy:}
				\begin{itemize}
					\setlength\itemsep{0em}
					\item Let $(A,B')$ and $(A', B)$ be edges present in $M_r$.
					\item $f(A,r) :=$ unique element in $A \cap B'$.
					\item $g(B,r) :=$ unique element in $A' \cap B$.
				\end{itemize}
				\caption{Rivest's strategy~\cite{Rivest}} \label{prot:rivest}
			\end{algorithm}
		\end{minipage}
	\end{figure}
	
	It is easy to see that both Alice's and Bob's outputs are uniformly distributed in $A$ and $B$, 
	respectively. Moreover, the probability that they output the same element, is exactly $1/a$, which is the probability of choosing the unique matching $M_r$ which contains the edge $(A,B)$ (\ie, enforcing $A' = A$ and $B' = B$).
	
	The strategy in the general case of $a \ne b$ is obtained by a simple reduction to the case above. Suppose w.l.o.g. that $a > b$. Alice and Bob get sets $A \subseteq [n]$ and $B \subseteq [n]$ such that $|A| = a$, $|B| = b$ and $|A \cap B| = 1$ and $A \cup B = [n]$. We extend the universe by adding $(a-b)$ dummy elements to get a universe of size $(2a-1)$ (note, $n = a+b-1$). Moreover, whenever Bob gets set $B$, he extends it to $B'$ by adding all the dummy elements to $B$ and thus $|B'| = a$ while having $|A \cap B'| = 1$ and $|A \cup B| = 2a-1$. Now, Alice and Bob can use the $\calG^{2a-1}_{a,a,1}$-restricted correlated sampling strategy from above on the input pair $(A, B')$. This achieves an error of $1 - 1/a = 1 - 1/\max\set{a, b}$. However, Bob's output is uniformly distributed over $B'$ and not $B$. To fix this, Bob can simply output a uniformly random element of $B$ whenever the above strategy requires him to return an element of $B' \setminus B$. It is easy to see that this doesn't change the error probability.
\end{proof}

\subsection{Proof of \autoref{thm:opt_large_inter}}\label{sec:large_int_pf}

\begin{proof}[Proof of \autoref{thm:opt_large_inter}]
	Let $A, B \subseteq [n]$ be such that $a = |A| = |B| = |A \cap B| + 1$ and let $P = \mathcal{U}(A)$ and $Q  = \mathcal{U}(B)$. For simplicity, we can assume without loss of generality that $A \cup B = [n]$. Thus, $n = a + 1$ and $\ell = a - 1$. Assume for the sake of contradiction that there is a $\calG^{a+1}_{a,a,a-1}$-correlated sampling strategy with disagreement probability $< 1 - \ell/(2a-\ell) = 2/n$. Let $\mathcal{D}$ be the uniform distribution over pairs $(A,B)$ of subsets of $[n]$ satisfying $A \cup B = [n]$ and $|A| = |B| = |A \cap B| + 1$. Note that $\calD$ is not a product distribution over $(A, B)$, unlike in \autoref{le:prod_opt}, which is what makes it challenging to analyze. By an averaging argument, there is a deterministic strategy pair $(f,g)$ such that,
	\begin{equation}\label{eq:cont}
		\Pr_{(A,B) \sim \mathcal{D}}[f(A) \neq g(B)] < \frac{2}{n}.
	\end{equation}
	Let 
	\begin{equation}
		i ~:=~ \argmax_{\ell \in [n]} \inabs{\set{A \in \binom{[n]}{n-1}: f(A)= \ell}}
	\end{equation}
	be the element that is most frequently output by Alice's strategy $f$, and denote its number of occurences by
	\begin{equation}
		k ~:=~ \bigg|\bigg\{A \in \binom{[n]}{n-1}: f(A)= i\bigg\}\bigg|.
	\end{equation}
	We consider three different cases depending on the value of $k$:
	\begin{enumerate} 
		\item[(i)] If $k \le n-3$, then consider
		a subset 
		$B \subseteq [n]$ with $|B| = n-1$. For any value of $f(B) \in B$, the conditional error probability $\Pr[f(A) \neq g(B)\, |\, B]$ is at least $2/(n-1)$. Averaging over all such $B$, we get a contradiction to \autoref{eq:cont}.
		\item[(ii)] If $k = n-2$, let $A_1 \neq A_2$ be the two subsets of $[n]$ with $|A_1| = |A_2| = n-1$ such that $f(A_1) \neq i$ and $f(A_2) \neq i$. For all $B \subseteq [n]$ with $|B| = n-1$ such that $B \neq A_1$ and $B \neq A_2$, the conditional error probability $\Pr[f(A) \neq g(B)\, |\, B]$ is at least $2/(n-1)$. Note that there are $n-2$ such
		sets $B$,   
		and that either $A_1$ or $A_2$ is the set $[n] \setminus \set{i}$. If $B = [n] \setminus \set{i}$, then the conditional disagreement probability $\Pr[f(A) \neq g(B)\, |\, B]$ is at least $(n-2)/(n-1)$. Averaging over all $B$, we get that
		\begin{equation*}
			\Pr_{(A,B) \sim \mathcal{D}}[f(A) \neq g(B)] \geq \inparen{\frac{2}{n-1}} \cdot \inparen{\frac{n-2}{n}} + \inparen{\frac{n-2}{n-1}} \cdot \inparen{\frac{1}{n}}  \ge \frac{2}{n}, 
		\end{equation*}
		where the last inequality holds for all $n \geq 2$. This contradicts \autoref{eq:cont}.
		\item[(iii)] If $k = n-1$, then the only subset $A_1$ of $[n]$ with $|A_1| = n-1$ and such that $f(A_1) \neq i$ is $A_1 = [n] \setminus \{i\}$. For all $B \neq A_1$, the conditional error probability $\Pr[f(A) \neq g(B)\, |\, B]$ is at least $1/(n-1)$. On the other hand, if $B = A_1$, then the conditional error probability is equal to $1$. Averaging over all $B$, we get that
		$$
		\Pr_{(A,B) \sim \mathcal{D}}[f(A) \neq g(B)] ~\geq~ \inparen{\frac{1}{n-1}} \cdot \inparen{ \frac{n-1}{n}} + 1 \cdot \inparen{\frac{1}{n}} ~=~ \frac{2}{n},
		$$
		which contradicts \autoref{eq:cont}.\qedhere
	\end{enumerate}
\end{proof}

\paragraph{Remark.} In~\cite{kleinberg2002approximation}, the correlated sampling strategy is used to give a randomized rounding procedure for a linear program. The factor $2$ loss in the correlated sampling strategy translates into an integrality gap of at most $2$. In fact, they also prove that the integrality gap is roughly tight. As pointed out by an anonymous reviewer, their proof essentially establishes \autoref{thm:opt_large_inter} under the assumption that $f = g$.

\section{Correlated sampling strategies of \cite{broder1997resemblance,kleinberg2002approximation,holenstein2007parallel}}\label{sec:protocols}

For sake of completeness, we describe the correlated sampling strategies of Broder and of Kleinberg--Tardos and Holenstein, thereby proving \autoref{thm:corr-samp}. 

\paragraph{Broder's Min Hash Strategy.} Consider the case of \emph{flat distributions}, where the distributions $P$ and $Q$ are promised to be of the following form: there exist $A, B \subseteq [n]$ such that $P = \mathcal{U}(A)$ and $Q  = \mathcal{U}(B)$. In this case, it is easy to show that the strategy given in \Cref{prot:minhash} achieves an error probability of $1 - |A \cap B|/|A \cup B|$. Since $\pi$ is a random permutation, $f(A,\pi)$ is uniformly distributed over $A$ and $g(B, \pi)$ is uniformly distributed over $B$. Let $i_0$ be the smallest index such that $\pi(i_0) \in A\cup B$. The probability that $\pi(i_0) \in A \cap B$ is exactly $|A \cap B|/|A \cup B|$, and this happens precisely when $f(A,\pi) = g(B,\pi)$. Hence, we get the claimed error probability.

\begin{figure}[ht]
	\centering
	\begin{minipage}{0.95\linewidth}
		\begin{algorithm}[H]
			{\bf Alice's input:} $A \subseteq [n]$\\[1mm]
			{\bf Bob's input:} $B \subseteq [n]$\\[1mm]
			{\bf Shared randomness:} a random permutation $\pi : [n] \to [n]$\\[2mm]
			{\bf Strategy:}
			\begin{itemize}  \setlength\itemsep{0em}
				\item $f(A,\pi) = \pi(i_A)$, where $i_A$ is the smallest index such that $\pi(i_A) \in A$.
				\item $g(B,\pi) = \pi(i_B)$, where $i_B$ is the smallest index such that $\pi(i_B) \in B$.
			\end{itemize}
			\caption{MinHash strategy~\cite{broder1997resemblance}}
			\label{prot:minhash}
		\end{algorithm}
	\end{minipage}
\end{figure}

The correlated sampling strategy of \cite{kleinberg2002approximation,holenstein2007parallel} follows a similar approach.

\begin{proof}[Proof of \autoref{thm:corr-samp}]
	Given a finite 
	set 
	$\Omega$ and probability
	distributions 
	$P$ and $Q$ over $\Omega$, define
	\[
	A := \setdef{(\omega, p) \in \Omega \times [0,1]}{p < P(\omega)} \quad\text{and}\quad B := \setdef{(\omega, q) \in \Omega \times [0,1]}{q < Q(\omega)}.
	\]
	Also for all $\omega \in \Omega$, define
	$A_\omega := A \cap (\set{\omega} \times [0,1])$ and
	$B_\omega := B \cap (\set{\omega} \times [0,1])$.
	
	The strategy of \cite{kleinberg2002approximation,holenstein2007parallel} can be intuitively understood as follows. 
	Alice and Bob use the MinHash strategy on inputs $A$ and $B$ over $\Omega \times [0,1]$, to obtain elements $(\omega_A, p_A)$ and $(\omega_B, p_B)$, respectively, 
	and simply output $\omega_A$ and $\omega_B$, respectively.
	However, this by itself is not well defined since $\Omega \times [0,1]$ is not a finite set. Nevertheless, the MinHash strategy can be modified to instead have a (countably) infinite sequence of points sampled i.\,i.\,d.\ from the uniform
	distribution 
	over $\Omega \times [0,1]$, instead of a permutation $\pi$. This strategy is summarized in \Cref{prot:holenstein}.
	
	Let $\mu$ be the uniform
	distribution 
	over $\Omega \times [0,1]$. Observe that $\mu(A) = \mu(B) = 1/|\Omega|$ and for all $\omega \in \Omega$, we have $\mu(A_{\omega}) = P(\omega)/|\Omega|$ and $\mu(B_{\omega}) = Q(\omega)/|\Omega|$. Similar to the analysis of the MinHash strategy, for Alice's chosen index $i_A$, we have $(\omega_{i_A}, r_{i_A})$ is uniform over $A$. Thus, $\Pr[f(P,\pi) = \omega]$ is precisely $\mu(A_{\omega}) / \mu(A) = P(\omega)$. Thus, $f(P,\pi)$ is distributed according to $P$ and similarly, $g(B,\pi)$ is distributed according to $Q$. Finally, $\Pr[f(P,\pi) = g(Q,\pi)] \ge \Pr[i_A=i_B]$. To bound this probability, note that $\mu(A\cap B) = (1-\delta)/|\Omega|$ and $\mu(A\cup B) = (1+\delta)/|\Omega|$.
	$$
	\Pr[f(P,\pi) = g(Q,\pi)]
	~\ge~ \Pr[i_A=i_B]
	~=~ \frac{\mu(A\cap B)}{\mu(A\cup B)}
	~=~ \frac{1 - \delta}{1+\delta}
	~=~ 1 - \frac{2\delta}{1+\delta}.
	$$
	\begin{figure}[ht]
		\centering
		\begin{minipage}{0.95\linewidth}
			\begin{algorithm}[H]
				{\bf Alice's input:} $P \in \Delta_\Omega$; let $A := \setdef{(\omega, p) \in \Omega \times [0,1]}{p < P(\omega)}$\\[1mm]
				{\bf Bob's input:} $Q \in \Delta_\Omega$; let $B := \setdef{(\omega, q) \in \Omega \times [0,1]}{q < Q(\omega)}$\\[1mm]
				{\bf Shared randomness:} An infinite sequence $\pi = ((\omega_1, r_1), (\omega_2, r_2), \ldots)$ where each $(\omega_i, r_i)$ is i.\,i.\,d.\ sampled uniformly from $\Omega \times [0,1]$.\\[2mm]
				{\bf Strategy:}
				\begin{itemize}  \setlength\itemsep{0em}
					\item $f(P,\pi) := \omega_{i_A}$, where $i_A$ is the smallest index such that $(\omega_{i_A}, r_{i_A}) \in A$
					\item $g(Q,\pi) := \omega_{i_B}$, where $i_B$ is the smallest index such that $(\omega_{i_B}, r_{i_B}) \in B$
				\end{itemize}
				\caption{The strategy of Kleinberg--Tardos and Holenstein~\cite{kleinberg2002approximation,holenstein2007parallel}} \label{prot:holenstein}   
			\end{algorithm}
		\end{minipage}
	\end{figure}
	We can ignore the possibility that no index $i_A$ exists satisfying $(\omega_{i_A},r_{i_A}) \in A$ (similarly for $B$) since this happens with probability $0$.
\end{proof}

\section{Discussion and open questions}\label{sec:discussions}

An immediate open question is to resolve \autoref{conj:rivest}. We reflect on some further open questions that are raised by the results discussed in this paper.

\subsection{Case of \emph{negatively correlated} sets}
In the context of \autoref{conj:rivest}, even in the setting where the set sizes are allowed to vary slightly, our knowledge is somewhat incomplete. \autoref{le:prod_opt} shows optimality of the MinHash strategy when $(A,B) \sim \calD_p$. In this case, $A$ and $B$ are independent and
each of them is $p$-biased,   
so $|A| \approx p \cdot n$, $|B| \approx p \cdot n$ and $|A \cap B| \approx p^2 \cdot n$. A simple reduction to \autoref{le:prod_opt} also implies the optimality of the MinHash strategy in the case where $A$ and $B$ are
\emph{positively correlated}.   
Specifically for parameters $\alpha > p$, consider the following distribution $\mathcal{D}_{p,\alpha}$ on pairs $(A,B)$ of subsets of $[n]$, where we first sample $S \subseteq [n]$ by independently including each element of $[n]$ with probability $p/\alpha$, and then independently including every $i \in S$ in each of $A$ and $B$ with probability $\alpha$. In this case, $|A| \approx p \cdot n$, $|B| \approx p \cdot n$ and $|A \cap B| \approx \alpha p \cdot n > p^2 \cdot n$. Even if we reveal $S$ to both Alice and Bob, \autoref{le:prod_opt} implies a lower bound of $2\delta/(1+\delta)$ on the error probability (for large enough $n$). It is unclear if the optimality holds even in the case where $A$ and $B$ are
\emph{negatively correlated},  
\ie, when $|A| \approx p \cdot n$, $|B| \approx p \cdot n$ and $|A \cap B| \ll p^2 \cdot n$.

\subsection{Fine-grained understanding of \texorpdfstring{$\calG$}{G}-restricted correlated sampling}
As alluded to in the 
Introduction,  
in the setting where $P$ and $Q$ are flat distributions on
subsets of $\Omega$ of different sizes,  
there is a strategy with lower error than provided in \autoref{thm:corr-samp}. In particular, for $P = \calU(A)$ and $Q = \calU(B)$ where $|A| \ne |B|$, the MinHash strategy gives an error probability of
\begin{equation}
	1 - \frac{|A \cap B|}{|A \cup B|}
\end{equation}
(which is the Jaccard distance between $A$ and $B$). However, na\"ively using the strategy of Kleinberg--Tardos and Holenstein would give an error probability of 
\begin{equation}
	1 - \frac{|A \cap B|}{|A \cup B| + ||A| - |B||},
\end{equation}  
which is higher than the Jaccard distance when $|A| \ne |B|$. This implies that the strategy of Kleinberg--Tardos and Holenstein is not
always optimal.  
Thus, it will be interesting to identify the right measure that captures the minimum error of a general $\calG$-restricted correlated sampling strategy.

\subsection{Correlated sampling for infinite spaces}
While this paper dealt with correlated sampling for finite sets $\Omega$, it might also be interesting to study it for infinite sets. This needs to be defined carefully in a measure theoretic sense, which could be done as follows. Consider a measure space $(\Omega, \calF, \mu)$, where $\Omega$ is the sample space, $\calF$ is a $\sigma$-algebra over $\Omega$ and $\mu$ is a finite measure on $(\Omega,\calF)$. Let $\Delta_{(\Omega, \calF, \mu)}$ be the set of all probability measures over $(\Omega,\calF)$ that are absolutely continuous with respect to $\mu$. The respective inputs of Alice and Bob are probability measures $P$ and $Q$ in $\Delta_{(\Omega, \calF, \mu)}$. A correlated sampling strategy for $(\Omega, \calF, \mu)$ is given by a pair of functions $f, g : \Delta_{(\Omega, \calF, \mu)} \times \calR\to \Omega$, where $f$ and $g$ are required to be measurable in their second argument $r \in \calR$. In order to define the error guarantee in terms of $\Pr_{r\sim\calR}[f(P,r)\ne g(Q,r)]$, however, we require that the event $\set{(\omega, \omega) : \omega \in \Omega}$ be measurable in $(\Omega \times \Omega, \calF \otimes \calF)$. This is true, for example, when $\Omega$ is a separable metric space equipped with the standard Borel algebra 
(see Chapters 3, 4 in~\cite{thorisson2000coupling}). We will assume this to be the case in the discussion henceforth and it might be useful to keep in mind a concrete example such as $\Omega = [0,1]$, equipped with the Lebesgue measure.

To the best of our knowledge, it remains open whether there exists a correlated sampling strategy for general measure spaces $(\Omega,\calF,\mu)$ with any \emph{non-trivial} error bound, that is, to get $\eps(\delta) < 1$ for all $\delta < 1$. This is in sharp contrast to coupling, where any two probability measures $P$ and $Q$ with $\dTV(P,Q) = \delta$ over $(\Omega, \calF)$ can be coupled with a disagreement probability of at most $\delta$.

Suppose the inputs $P$ and $Q$ are promised to be such that the corresponding Radon--Nikodym derivatives (a.\,k.\,a.\ \emph{densities}) $dP/d\mu$ and $dQ/d\mu$ 
are bounded everywhere by a known constant $c$. Then it is possible to generalize the strategy of Kleinberg--Tardos and Holenstein (\Cref{prot:holenstein}) and get the same error guarantee as in \autoref{thm:corr-samp}; this can be done by using $\mu$ instead of the uniform measure on $\Omega$ and replacing $[0,1]$ by $[0,c]$.

However, the problem gets challenging if there is no promised upper bound on the Radon--Nikodym derivatives. One explanation for why this challenge is not faced in obtaining a coupling is because knowing 
both $P$ and $Q$,    
we can always take $\mu' = (P+Q)/2$ as a measure with respect to which both $P$ and $Q$ are absolutely continuous and more strongly, the Radon--Nikodym derivatives $dP/d\mu'$ and $dQ/d\mu'$ are
never greater than $2$.    
On the other hand, for correlated sampling, the players do not have access to such a common $\mu'$.

It might also be interesting to study a generalized notion of error in correlated sampling strategies where we wish to minimize $\Ex_{r\sim\calR} [d(f(P,r), g(Q,r))]$ for some metric $d : \Omega \times \Omega \to \bbR_{\ge 0}$ over $\Omega$. The error guarantee studied in this paper corresponds to the discrete metric $d(x,y) = \mathbbm{1}\set{x\ne y}$. For $\Omega \subseteq \bbR$, such as $\Omega = [0,1]$, we might alternatively want to consider $d(x,y) = |x-y|$. Since a correlated sampling strategy induces a coupling, this notion of error can never be lower than the Wasserstein distance $W_1(P,Q)$ (also known as Earth-Mover distance) between the distributions $P$ and $Q$. To the best of our knowledge, it remains open in this setting, whether correlated sampling strategies can get an error that is never larger than some function of $W_1(P,Q)$.

\bibliographystyle{alpha}
\bibliography{correlated_sampling.bbl}

\end{document}